\documentclass[twocolumn]{aastex7}

\usepackage{caption}
\usepackage{subcaption}
\usepackage{booktabs}
\usepackage{siunitx}
\usepackage{multirow}
\usepackage{bm}

\begin{document}
			
\title[Orbital Stability of Apsidally Aligned Planetary Systems]{Orbital Stability of Apsidally Aligned Planetary Systems}

\author[orcid=0000-0003-1890-6356,gname=Gabriel, sname='Teixeira Guimar\~es']{Gabriel Teixeira Guimar\~aes}
\affiliation{Department of Astronomical Science, The Graduate University for Advanced Studies, SOKENDAI, 2-21-1 Osawa, Mitaka, 181-8588, Tokyo, Japan.}
\affiliation{Division of Science, National Astronomical Observatory of Japan, NAOJ, 2-21-1 Osawa, Mitaka, 181-8588, Tokyo, Japan.}
\email[show]{gabriel.guimaraes@grad.nao.ac.jp}  
			
\author[orcid=0000-0002-5486-7828,gname=Eiichiro, sname='Kokubo']{Eiichiro Kokubo} 
\affiliation{Department of Astronomical Science, The Graduate University for Advanced Studies, SOKENDAI, 2-21-1 Osawa, Mitaka, 181-8588, Tokyo, Japan.}
\affiliation{Division of Science, National Astronomical Observatory of Japan, NAOJ, 2-21-1 Osawa, Mitaka, 181-8588, Tokyo, Japan.}
\affiliation{Center for Computational Astrophysics, CfCA, National Astronomical Observatory of Japan, NAOJ, 2-21-1 Osawa, Mitaka, 181-8588, Tokyo, Japan}
\email{kokubo.eiichiro@nao.ac.jp}
			
			
\begin{abstract}

The dynamical stability of compact planetary systems is known to depend strongly on orbital eccentricities, but the physical role of apsidal alignment remains poorly understood. 
We investigate this problem using direct {\em N}-body simulations by separating planetary eccentricities into a systematic component shared by all planets and a random component that determines the degree of apsidal alignment.
We find that these two components play fundamentally different dynamical roles. 
Random eccentricity destabilizes planetary systems by driving neighboring planet pairs into the regime of mean-motion resonance overlap, leading to chaotic evolution.
In contrast, systematic eccentricity has little effect on the instability timescale over a wide range of values because the planetary orbits remain in apsidal alignment, preserving nearly fixed relative orbital orientations despite finite eccentricities. 
As a result, systems that start in apsidal alignment are nearly as stable as initially circular systems.
This protection mechanism has a finite domain of validity.
Beyond a critical systematic eccentricity, coherent apsidal precession breaks down, differential precession develops, and the instability timescale decreases rapidly.
Our results demonstrate that the long-term stability of compact planetary systems cannot be characterized solely by orbital eccentricities but also depends on the maintenance of coherent apsidal precession, providing a new direction for analytical stability criteria.

\end{abstract}
			
		
		

\section{Introduction}\label{sec1}

The Kepler space telescope discovered more than 2000 planetary systems (\citealt{exoplanet}), many of which have planets on short-period orbits ($P<1$ yr).
Around $20\%$ of those contain more than two planets (henceforth ``Kepler multis''), and are commonly found in compact orbits with small orbital period ratio $P_{i+1}/P_{i}<2$.  They have similar orbital spacing between adjacent planets (measured by the mutual Hill radius $R_\mathrm{H}^{i,i+1}$, henceforth just called $R_\mathrm{H}$), and similar radii.
Such systems are called ``peas-in-a-pod'' (\citealt{weiss_peas}) and most orbit billion-year-old sun-like stars, leading to the belief that those systems are dynamically stable.

On the other hand, theory of planet formation indicates that the phase of oligarchic growth naturally leads to protoplanet systems with similar masses and Hill-radius scaled orbital separations $K$ ($a_{i+1}-a_{i}\approx KR_\mathrm{H}$, e.g. \citealt{ida_kokubo, kambara25}). 
Their subsequent evolution is driven by dynamical instabilities and giant impacts, which reshape the system's dynamical architecture. 

The onset of instabilities was first numerically explored by \cite{chambers_stability_1996}, who found a log-linear relation between the close encounter timescale and the initial orbital separation $\log_{10}\tau_{\mathrm{inst}}=bK+c$, for circular coplanar systems. 
Studies by \cite{yoshinaga_stability_1999, zhou_2007} showed that small eccentricities can reduce $\tau_{\mathrm{inst}}$ by up to two orders of magnitude. 
\cite{gratia} observed that this dependency is influenced by the apsidal configuration: eccentric systems in apsidal alignment ($\varpi_{i+1} = \varpi_{i}$) can survive for as long as circular systems.
\cite{pu_spacing_2015} observed that the orbital distribution of Kepler Multis is consistent with planets forming farther from their stars or evolving from initially more compact configurations. 

\cite{quillen, petit_path_2020} identified the overlap of neighboring three-body mean-motion resonances (3BMMRs) as the source of instabilities in circular compact systems and derived estimates for the instability timescales. 
For eccentric systems, \cite{lithwick} extended the resonance-overlap criterion of \cite{wisdom_overlap} by accounting for higher-order two-body MMRs, deriving a critical eccentricity $e_{\mathrm{crit}}$ for the onset of resonance overlap, and showed that it can be written in terms of the relative eccentricity $\bm{e}_{i,i+1}=\bm{e}_i-\bm{e}_{i+1}$ rather than the individual eccentricities. 
\cite{tamayo21} generalized those to multi-planet systems, showing that equally eccentric, apsidally aligned planet pairs have $\bm{e}_{i,i+1}=0$, and are expected to be more stable than systems with randomly distributed apsidal angles, consistent with the numerical results of \cite{gratia}.

Some questions remain unanswered, though: What is the physical mechanism behind the enhanced stability in apsidally aligned systems?
What is the limitation of the dynamical protection provided by apsidal alignment?
Under what conditions does it break down?
		
Motivated by those questions, and by the discovery of apsidally aligned extrasolar planets (\citealt{trifonov1, laune, brahm_aligned}), we conducted a high-resolution numerical study of the sensitivity of instability timescales $\tau_{\mathrm{inst}}$ to initial eccentricities and longitudes of pericenter, to better understand the changes in stability thresholds from systems in apsidal alignment to ones with randomized pericenters. 

To achieve this, we decompose planetary eccentricities into systematic and dispersion components, and use this framework to identify the physical mechanism responsible for the enhanced stability of apsidally aligned planetary systems and the conditions under which this protection breaks down.

The structure of this paper is as follows:
In Section \ref{sec2}, we describe our numerical methods and initial conditions for numerical integration.
In Section \ref{sec3}, we show the results obtained for the effects of initial eccentricities and longitudes of pericenter on the instability timescales.
We also present the conditions for apsidal alignment protection and their limitations.
In Section \ref{sec4}, we discuss our results in the context of previous studies, and present our conclusions in Section \ref{sec5}.

\section{Calculation Method}\label{sec2}

\subsection{Instability definitions}\label{sec221}

Many definitions of ``instability'' were used in the literature
\citep{chambers_stability_1996, yoshinaga_stability_1999, bartram, wang_influence_2019, doty_25} as means of better characterizing instability timescales and to allow for economy of computational resources by stopping simulations earlier than the total integration time.
		
To analyze the effect of the choice of the definition of instability on the instability timescales, for each set of parameters integrated, four definitions are considered:
\begin{enumerate}
    \item Close encounter time, $\tau_{\mathrm{close}}$: $r_{ij}<R_\mathrm{H}$;
    \item Orbital flip time, $\tau_{\mathrm{flip}}$: $a_{i}>a_{i+1}$;
    \item Orbital crossing time, $\tau_{\mathrm{cross}}$: $a_{i}(1+e_{i})>a_{i+1}(1-e_{i+1})$;
    \item Physical collision time, $\tau_{\mathrm{coll}}$: $r_{ij}<R_\mathrm{p}^{i}+R_\mathrm{p}^{j}$,
\end{enumerate}
where $r_{ij}$ is the distance between planets $i$ and $j$, and
\begin{equation}
	R_\mathrm{H} = \frac{a_{i}+a_{j}}{2}\left(\frac{m_{i}+m_{j}}{3M_{\ast}}\right)^{1/3}
\end{equation} 
is the mutual Hill radius between them. 

For all simulations conducted throughout this paper, we consider $\tau_\mathrm{inst} = \tau_\mathrm{close}$, and justify our choice in Appendix \ref{sectrig}.
		
\subsection{Initial conditions}\label{sec22}

All integrated systems consist of $N=5$ Earth-mass planets orbiting a Solar-mass star. 
The innermost planet has a fixed semimajor axis $a_{1} = 0.1$ au ($P_1\approx 11.5$ days). 
		
The orbital separation between pairs of adjacent planets is drawn so as to keep $a_{i+1} - a_{i} = KR_\mathrm{H}$. 
All planet pairs have the same $K$, meaning these are equal-mass and spacing (EMS) systems.
The true anomaly of all planets is randomly drawn as $f\in[0, 360^{o})$.

\begin{figure}[!ht]
    \centering\resizebox{1\hsize}{!}{\includegraphics[width=0.25\textwidth]{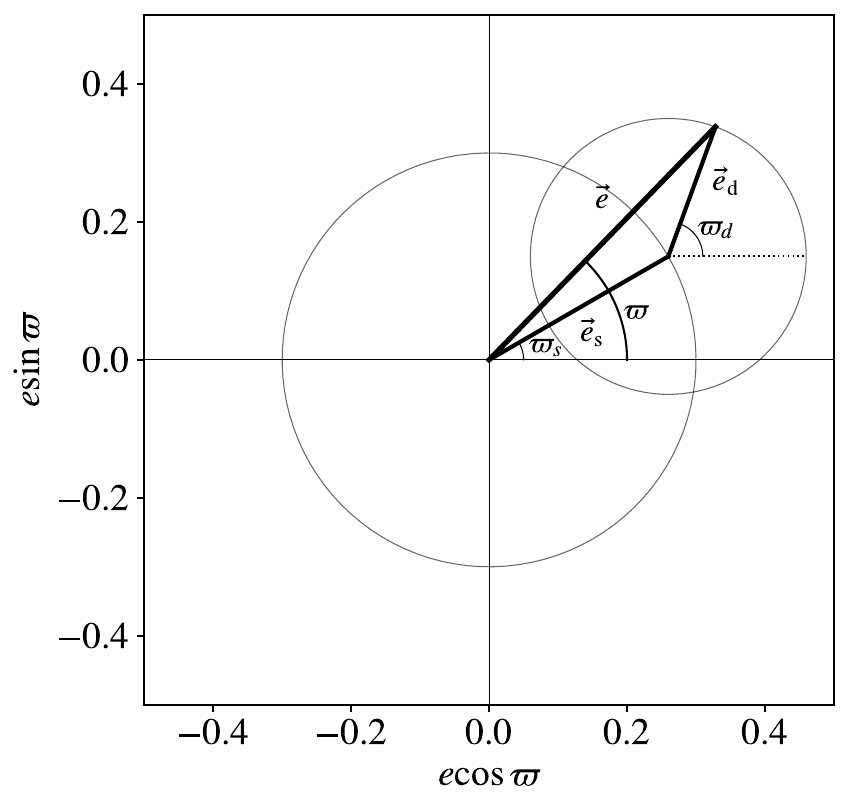}}
    \caption{\footnotesize Schematic diagram of how total planetary eccentricity vectors $\vec{e}$ are decomposed as the sum of a systematic eccentricity vector $\vec{e}_\mathrm{s}$ and an eccentricity dispersion vector $\vec{e}_\mathrm{d}$, $\vec{e} = \vec{e}_\mathrm{s} + \vec{e}_\mathrm{d}$. The diagram was constructed with initial conditions $(e_\mathrm{s}, \varpi_\mathrm{s}) = (0.2, 30^{o})$ and $(e_\mathrm{d}, \varpi_\mathrm{d}) = (0.15, 70^{o})$.}
    \label{free_forc}
\end{figure}

Previous works on eccentric systems relied on randomized longitudes of pericenter $\varpi_{i}$.
In this work, we decompose each planet's eccentricity vector $\bm{e}$ in two components: systematic eccentricity $\bm{e}_{\mathrm{s}}$ and dispersion eccentricity $\bm{e}_{\mathrm{d}}$, from which we define the total eccentricity vector $\bm{e} = \bm{e}_{\mathrm{s}}+\bm{e}_{\mathrm{d}}$. 
The systematic and dispersion eccentricity vectors are defined as $\bm{e}_{\mathrm{s}} = (e_{\mathrm{s}}\cos\varpi_{\mathrm{s}}, e_{\mathrm{s}}\sin\varpi_{\mathrm{s}})$ and $\bm{e}_{\mathrm{d}} = (e_{\mathrm{d}}\cos\varpi_{\mathrm{d}}, e_{\mathrm{d}}\sin\varpi_{\mathrm{d}})$, where $\varpi_{\mathrm{s}}$ and $\varpi_{\mathrm{d}}$ are the systematic and dispersion longitudes of pericenter, respectively.
A schematic diagram of this vectorial decomposition can be seen in Figure \ref{free_forc}.

For each planetary system, the length and orientation of $\bm{e}_{\mathrm{s}}$ are fixed for all planets, with $\varpi_{\mathrm{s}} = 0$ and $e_{\mathrm{s}}$ as a free parameter for initializing the simulations. 
For $\bm{e}_{\mathrm{d}}$, $e_{\mathrm{d}}$ is drawn from a Rayleigh distribution with scale parameter $\sigma_{\mathrm{e}}$ 
\begin{equation}
	f(e_{\mathrm{d}}, \sigma_{\mathrm{e}}) = \frac{e_{\mathrm{d}}}{\sigma_{\mathrm{e}}^{2}}\exp(-e_{\mathrm{d}}^{2}/2\sigma_{\mathrm{e}}^{2}),
\end{equation} 
while $\varpi_{\mathrm{d}}$ are randomly drawn between $\left[0, 360^{o}\right)$. 
The scale parameter $\sigma_{\mathrm{e}}$ is also a free parameter.

For $e_{\mathrm{d}}\ll e_{\mathrm{s}} \longrightarrow\bm{e}\approx\bm{e}_{\textbf{s}}$, in which case the planetary systems have aligned longitudes of pericenter and relative eccentricity $\bm{e}_{i}-\bm{e}_{j}\propto\sigma_{\mathrm{e}}$. 
On the other hand, for $e_{\mathrm{s}}\ll e_{\mathrm{d}}$, pericenter directions can be considered random.

The values $e_{\mathrm{s}}$ and $\sigma_{\mathrm{e}}$ are drawn logarithmically from the interval $[10^{-4}, 10^{-1}]$ and linearly from the interval $[0.1, 0.5]$, except for $e_{\mathrm{s}} = 0$ and $\sigma_{\mathrm{e}} = 0$. 
The range of $\sigma_{\mathrm{e}}$ is based on the distribution of observed eccentricities in the Kepler multis ($\langle e\rangle\approx0.03$), while $e_{\mathrm{s}}$ was extended to possibly span stable and unstable configurations.

Two sets of runs are conducted: a high-resolution survey of 801 evenly sampled $K$ in which all sets of parameters $(K, e_{\mathrm{s}}, \sigma_{\mathrm{e}})$ are integrated once, and a low-resolution set of 9 $K$ values, in which all sets of parameters $(K, e_{\mathrm{s}}, \sigma_{\mathrm{e}})$ are integrated 20 times. 
The high-resolution survey is designed to resolve the fine structure of the phase space, whereas the ensemble runs are intended to quantify the statistical robustness of the instability timescales. 
As demonstrated in Appendix \ref{apdxens}, the results obtained from the high-resolution single-run survey
are statistically consistent with those derived from ensemble integrations and do not affect the results presented throughout the paper.

The values of $K$ are chosen based on previous numerical experiments that reproduce both stable and unstable regimes for a broad range of architectures.

For the survey on instability definitions described in Subsection \ref{sec221}, a specific set of initial conditions is used: 801 initial orbital separations $K\in[4,10]$ are integrated for different sets of initial eccentricities $e = 0, 0.02, 0.05$ and mutual inclinations $I = 0^{o}, 1^{o}, 10^{o}$, in which all planets are assigned the same eccentricities and the same mutual inclination between adjacent pairs of planets.
The longitudes of pericenter $\varpi$ and longitudes of ascending node $\Omega$ are  randomly drawn from $[0, 360^{o})$.
The results are shown in Appendix~\ref{sectrig}.

For the collisional trigger, planetary radius $R_\mathrm{p}$ is calculated considering an Earth-like density ($\rho_\oplus \approx 10^{7}M_{\oplus}/\mathrm{au}^{3}, M_\oplus \approx 3\times 10^{-6}M_\odot$) for all planets.

All initial conditions are integrated for $10^{8} P_{1}$ or until the instability definitions are met.

A summary of all initial conditions used is shown in Table \ref{sum_ic}.
		
		\begin{table}[ht!]
			\centering
			\caption{Summary of the initial conditions.}
			\begin{tabular}{lc}
				\hline
				\textbf{Parameter} & \textbf{Value / Range} \\
				\hline
				$M_\ast$ & $1\,M_\odot$ \\
				$m_\mathrm{p}$ & $1\,M_\oplus$ \\
				$N$ & 5 \\
				$a_1$ & $0.1\,\mathrm{au}$ \\
				$P_1$ & $\simeq 11.55$ days \\
				$f$ & Random in $[0, 360^{o})$ \\
				$e_{\mathrm{s}}$ & $\in[0.0,0.5]$ \\
				$\sigma_{\mathrm{e}}$ & $\in[0.0,0.1]$ \\
				$\varpi_{\mathrm{s}}$ & $0$ \\
				$\varpi_{\mathrm{d}}$ & Random in $[0, 360^{o})$ \\
				$I$ & $0^{o}, 1^{o}, 10^{o}$ \\
                $K$ & Evenly spaced in [4, 20] \\
				\hline
			\end{tabular}
			\label{sum_ic}
		\end{table}

\subsection{Orbit integration}\label{sec21}
        
All integrations are conducted using the REBOUND {\em N}-body code \citep{Rein_2015}. 
The hybrid integrator TRACE is chosen for its accuracy in handling close encounters and the efficiency of a symplectic integrator \citep{trace}. 
The minimum timestep used during WHfast periods is set to $\Delta t = 0.05P_{1}$. 
        
\section{Results}\label{sec3}

\subsection{Role of eccentricity dispersion}\label{sec32}

\begin{figure*}[!ht]
    \centering\resizebox{.95\hsize}{!}{\includegraphics[width=0.975\textwidth]{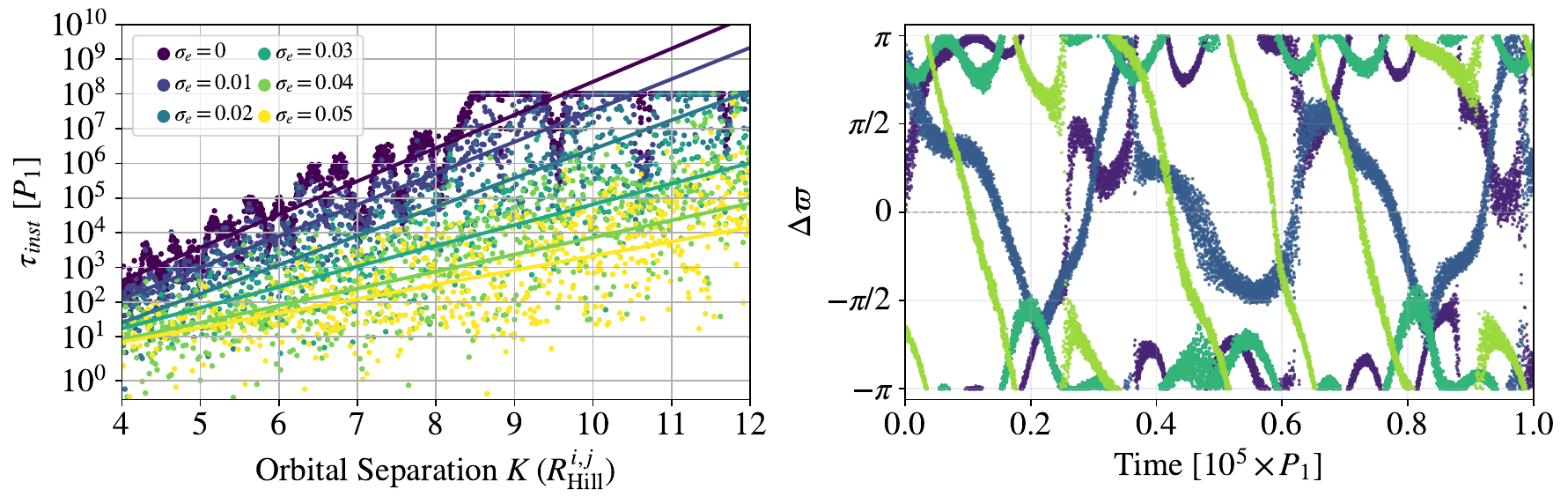}}
    \caption{Panel a): Instability timescales $\tau_{\mathrm{inst}}$ as a function of orbital separation $K$ for initially randomized longitudes of pericenter ($e_{\mathrm{s}} = 0$) for different initial eccentricities. Straight lines are log-linear fits $\log_{10}\tau_{\mathrm{inst}} = bK+c$. Panel b): Time evolution of the relative apsidal angle $\Delta\varpi_{i, i+1} = \varpi_{i+1}-\varpi_{i}$ for an orbit with $K = 10, \sigma_{\mathrm{e}} = 0.01$. Different colors refer to $\Delta\varpi$ between all adjacent pairs of planets.}
    \label{dense_tinst_ks_ed}
\end{figure*}
        
The instability timescales $\tau_{\mathrm{inst}}$ are characterized as a function of initial orbital separation $K$ for the cases with randomized longitudes of pericenter ($e_{\mathrm{s}}  = 0$). 
Panel a) of Figure \ref{dense_tinst_ks_ed} shows $\tau_{\mathrm{inst}}$ over a dense grid of points in $K$ for $\sigma_{\mathrm{e}}\in[0, 0.01, 0.02, 0.05]$. 
In panel b), an example of the time evolution of $\Delta\varpi$ of the four pairs of planets for a given initial condition is shown (see caption for details).
		
The effects of $\sigma_{\mathrm{e}}$ are noticeable, as circular systems can reach $\tau_\mathrm{inst} = 10^{8}P_{1}$ at $K\approx8.5$ whereas $\tau_{\mathrm{inst}}$ roughly decreases one order of magnitude in systems with $\sigma_{\mathrm{e}} = 0.01$ for the same orbital separation. 
As $\sigma_{\mathrm{e}}$ increases, $\tau_\mathrm{inst}$ decreases substantially, completely destroying the resonant structure.
For $\sigma_{\mathrm{e}} = 0.05$, none of the systems survived the whole integration time.

Increasing $\sigma_{\mathrm{e}}$ can put planet pairs in the resonance overlap region, which can take place for substantially lower eccentricities in compact planetary systems, leading to chaotic dynamical evolution \citep{wisdom_overlap, lithwick}.
Once their orbits begin to precess differentially (panel b of Figure~\ref{dense_tinst_ks_ed}), sufficiently large eccentricities allow neighboring orbits to intersect, sharply decreasing their survival times.

\begin{figure}
    \centering\resizebox{1.\hsize}{!}{\includegraphics[width=0.975\textwidth]{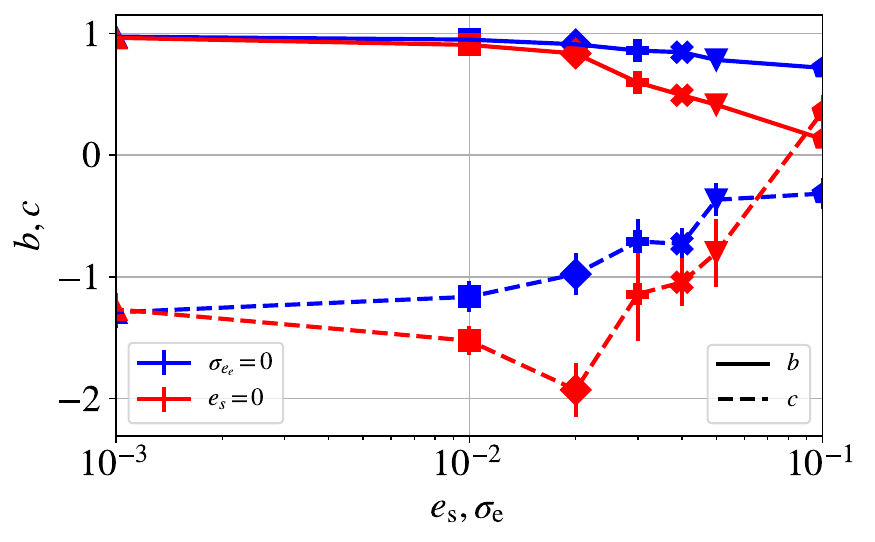}}
    \caption{Coefficients $b$ and $c$ obtained for the log-linear fit $\log_{10}\tau_{\mathrm{inst}} = bK+c$ as a function of $e_{\mathrm{s}}$ and $\sigma_{\mathrm{e}}$. Blue points denote fits in apsidally aligned systems, while red points denote fits in randomized pericenter-longitude systems. Solid curves refer to coefficient $b$, dashed ones refer to coefficient $c$. For completeness, the plot includes the coefficients obtained for additional $e_\mathrm{s},\sigma_\mathrm{e} = 0.001, 0.03, 0.04, 0.1$.}
    \label{coefsbc}
\end{figure}

For each initial $\sigma_{\mathrm{e}}$, a log-linear fit of the form $\log_{10}\tau_{\mathrm{inst}} = bK+c$ is applied considering $K\leq 8.5$ due to the saturated values of $\tau_{\mathrm{inst}}$.
The coefficients $b$ and $c$ and their respective error bars are listed in Table \ref{tab_fit_es}, while a plot of $b$ and $c$ as a function of $\sigma_{\mathrm{e}}$ is shown in Figure \ref{coefsbc}.

For $\sigma_{\mathrm{e}}\leq 0.01$, there is little to no spread in $b$ and $c$. 
For $\sigma_{\mathrm{e}} \geq 0.03$, $b$ decreases drastically, while $c$ increases. 

Comparing our fits with results in the literature, for $\sigma_e = 0.01$, we obtained $b = 0.905\pm0.018$, whereas \cite{yoshinaga_stability_1999} obtained a $b = 0.623\pm0.06$.
This discrepancy is due to their non-zero initial inclination, leading to larger $\tau_{\mathrm{inst}}$ for smaller $K$, resulting in a decrease in slope $b$, as observed in Figure \ref{tinst_ks_bench_triggers}.

For $\sigma_e = 0.05$, \cite{gratia} obtained $b = 0.479\pm0.006$, while we obtained $b = 0.413\pm0.041$. 
Although small, the difference between the two results is not within $1\sigma$. 
We attribute this discrepancy to a difference in the definition of close encounters: we consider a close encounter to happen whenever $r_{ij} \leq R_\mathrm{H}$ (i.e., the close approach distance will depend on the semi-major axes of the planets and their masses through $R_\mathrm{H}$), whereas they fix the close encounter distance as $r_{ij}<0.01$ au. 
In this case, our simulation tends to detect close encounters earlier than \cite{gratia} for larger $K$. This is due to the fact that our close encounter radius increases with the semi-major axis of the outer planets, leading to lower $b$, whereas theirs keeps it constant for all $K$.

\begin{table}[ht!]
    \centering
    \caption{Log-linear fit coefficients for different $\sigma_{\mathrm{e}}$ ($e_{\mathrm{s}} = 0$)}
    \begin{tabular}{cccccc}
        \toprule
        $\sigma_{\mathrm{e}}$ & $b$ & $\sigma_{b}$ & $c$ & $\sigma_{c}$ \\
        \midrule
       $0$     & 0.956 & 0.015 & -1.203 & 0.095 \\
       $0.01$  & 0.905 & 0.019 & -1.527 & 0.119 \\
       $0.02$  & 0.834 & 0.035 & -1.930 & 0.220 \\
       $0.05$  & 0.413 & 0.041 & -0.803 & 0.280 \\
        \bottomrule
    \end{tabular}
    \label{tab_fit_es}
\end{table}
        
\subsection{Role of systematic eccentricity}\label{sec33}

\begin{figure*}[!ht]
    \centering\resizebox{.95\hsize}{!}{\includegraphics[width=0.975\textwidth]{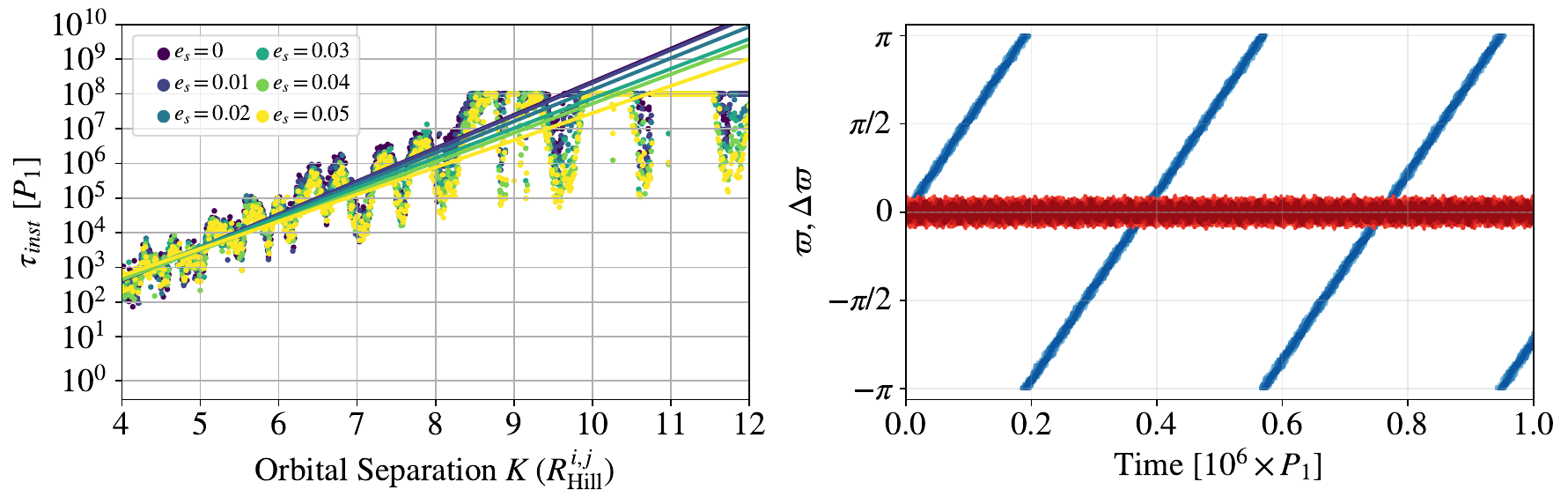}}
    \caption{Panel a): Instability timescales $\tau_{\mathrm{inst}}$ as a function of orbital separation $K$ forapsidally aligned systems ($\sigma_{\mathrm{e}} = 0$) for different initial systematic eccentricities. Straight lines are log-linear fits $\log_{10}\tau_{\mathrm{inst}} = bK+c$. . Panel b) shows the time evolution of $\varpi$ and $\Delta\varpi$ (blue and red lines, respectively) for all the planets in a system with $K = 10, e_\mathrm{s} = 0.01$.}
    \label{dense_tinst_ks_es}
\end{figure*}

In contrast to the effects of $\sigma_\mathrm{e}$, increasing systematic eccentricity has limited effects on $\tau_\mathrm{inst}$ for a wide range of $e_\mathrm{s}$. 
The resonant structure remains clear for all $e_\mathrm{s}$, and $\tau_{\mathrm{inst}}$ quickly saturates at the integration time $\tau_{\mathrm{inst}} = 10^{8}P_{1}$ at the same $K$ for all $e_\mathrm{s}$.
The only noticeable difference lies in the \textit{loci} of MMR, which gets broader for $e_\mathrm{s}=0.05$.

Panel b) of Figure \ref{dense_tinst_ks_es} shows the mechanism allowing for aligned systems to be long-lived: the longitudes of pericenter $\varpi_{i}$ of all planets circulate altogether, with the same precession frequency, leading all $\Delta\varpi_{i,i+1}$ to oscillate around 0 with a small amplitude throughout the whole integration.
		
The rigid-body-like precession provides dynamical protection: as long as this coherent precession persists, neighbouring orbits maintain nearly constant relative orientations, suppressing orbit crossings despite their finite eccentricities. 
This enhances the stability of such systems and allows them to outlive systems in which no apsidal alignment is observed.
		
The stability of aligned systems is evident from the log-linear fits, since the slope $b$ is almost unchanged for a broad range of $e_\mathrm{s}$.
The coefficients obtained for the randomized case quickly diverge from the aligned case as $\sigma_\mathrm{e}$ increases.

\begin{table}[ht!]
    \centering
    \caption{Log-linear fit coefficients for different $e_s$ ($\sigma_{\mathrm{e}} = 0$)}
    \begin{tabular}{cccccc}
        \toprule
        $e_s$ & $b$ & $\sigma_{b}$ & $c$ & $\sigma_{c}$ \\
        \midrule
        $0.0$ & 0.956 & 0.015 & -1.203 & 0.095 \\
        $0.01$ & 0.948 & 0.020 & -1.166 & 0.128 \\
        $0.02$ & 0.910 & 0.027 & -0.979 & 0.175 \\
        $0.05$ & 0.782 & 0.021 & -0.368 & 0.136 \\
        \bottomrule
    \end{tabular}
    \label{tab_fit_ed}
\end{table}
        		
Since only \cite{gratia} considered the effects of apsidal alignment, we therefore compare our results for $e_{\mathrm{s}} = 0.05$ with theirs. 
While they obtained $b = 0.837\pm0.016$, we obtained $b = 0.781\pm0.022$. 
This discrepancy can be explained by noting that they integrated their systems for $100\times$ longer, up to $10^{10}P_{1}$. 
The prolonged integration times ``shift'' the $\tau_{\mathrm{inst}}$ saturation to larger $K$, where stronger MMR lie, and have a further impact on survival times, leading to larger $b$. 
When considering only $K\leq8$ (where stronger MMR do not lead to drastic oscillations in $\tau_{\mathrm{inst}}$), we obtained $b = 0.827\pm0.023$, consistent with their results.

In light of these results, one could ask: How eccentric can apsidally aligned planetary systems be before becoming unstable? 
In which eccentricity range does the phase protection provided by the apsidal alignment hold? 
We study these subjects in more detail in the next subsection.

\subsection{Mixed configurations}{\label{sec34}}

\begin{figure*}[!ht]
    \centering\resizebox{1.\hsize}{!}{\includegraphics[width=0.975\textwidth]{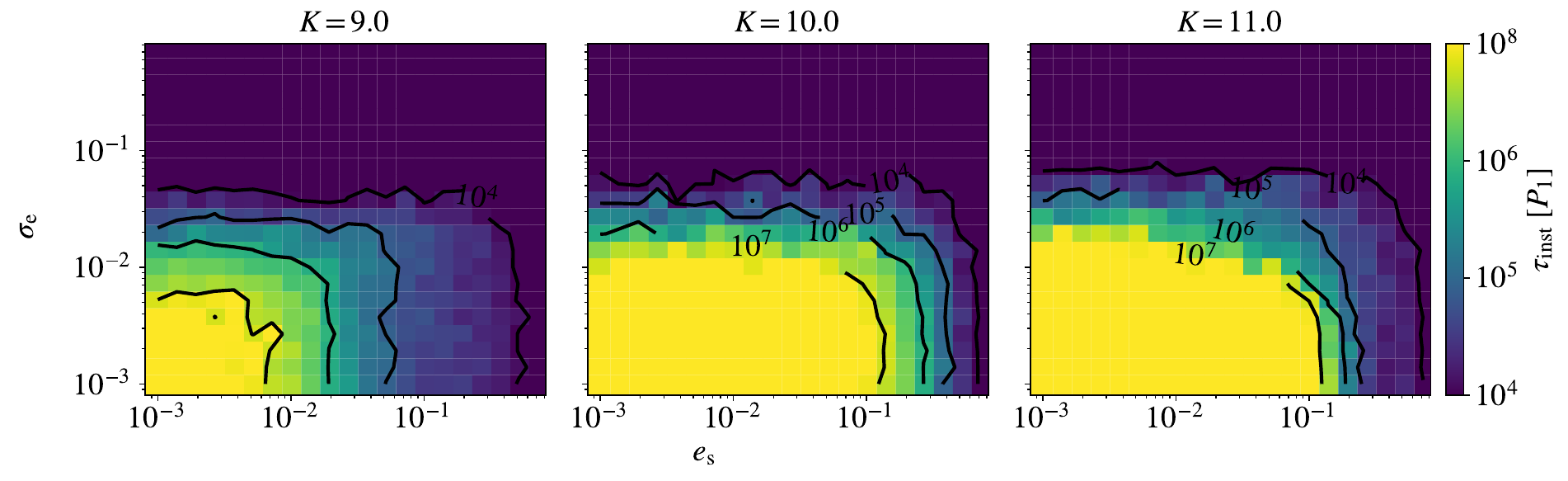}}
    \caption{Stability maps of the instability timescale $\tau_{\mathrm{inst}}$ as a function of systematic and dispersion eccentricities $e_{\mathrm{s}}, \sigma_{\mathrm{e}}$ for different orbital separations $K = 8, 9, 10$. Lighter colors indicate more stable regions, whereas darker colors indicate shorter $\tau_{\mathrm{inst}}$. Black contours indicate $\tau_{\mathrm{inst}}\in[10^{4}, 10^{5}, 10^{6}, 10^{7}]P_{1}$.}
    \label{tinst_map_es_ed}
\end{figure*}

Up to this point, the systematic and dispersion components have been investigated separately. However, real planetary systems generally contain both components simultaneously. 

We now mix the two non-zero components of our eccentricity vector, $\bm{e}_{\mathrm{s}}$ and $\bm{e}_{\mathrm{d}}$, to observe the effects of mixing both eccentricities on the dynamical stability of planetary systems.

We first construct a map of $\tau_{\mathrm{inst}}$ as a function of initial $e_{\mathrm{s}}$ and $\sigma_{\mathrm{e}}$ for three different initial $K$. 
The maps are shown in Figure \ref{tinst_map_es_ed}.

All maps display the lower-left region of enhanced stability related to low-eccentricity, quasi-circular orbits.
As the orbital separation increases, the stability region expands asymmetrically towards larger eccentricities. 
The asymmetry is biased towards larger $e_{\mathrm{s}}$, while $\tau_\mathrm{inst}$ quickly decreases as $\sigma_{\mathrm{e}}$ gets larger. 
The exception is the map for $K = 9$, which corresponds to a region of MMR, where a dip in $\tau_{\mathrm{inst}}$ is visible in Figures \ref{dense_tinst_ks_ed} and \ref{dense_tinst_ks_es}.

The stability maps show that fundamentally different dynamical roles are at play. 
They also show that the apsidal protection mechanism has its own limitations and cannot be maintained at sufficiently large $e_\mathrm{s}$, motivating further investigation of this regime.

\subsubsection{Effects of apsidal alignment}\label{sec331}

\begin{figure*}
    \centering\resizebox{1.\hsize}{!}{\includegraphics[width=0.975\textwidth]{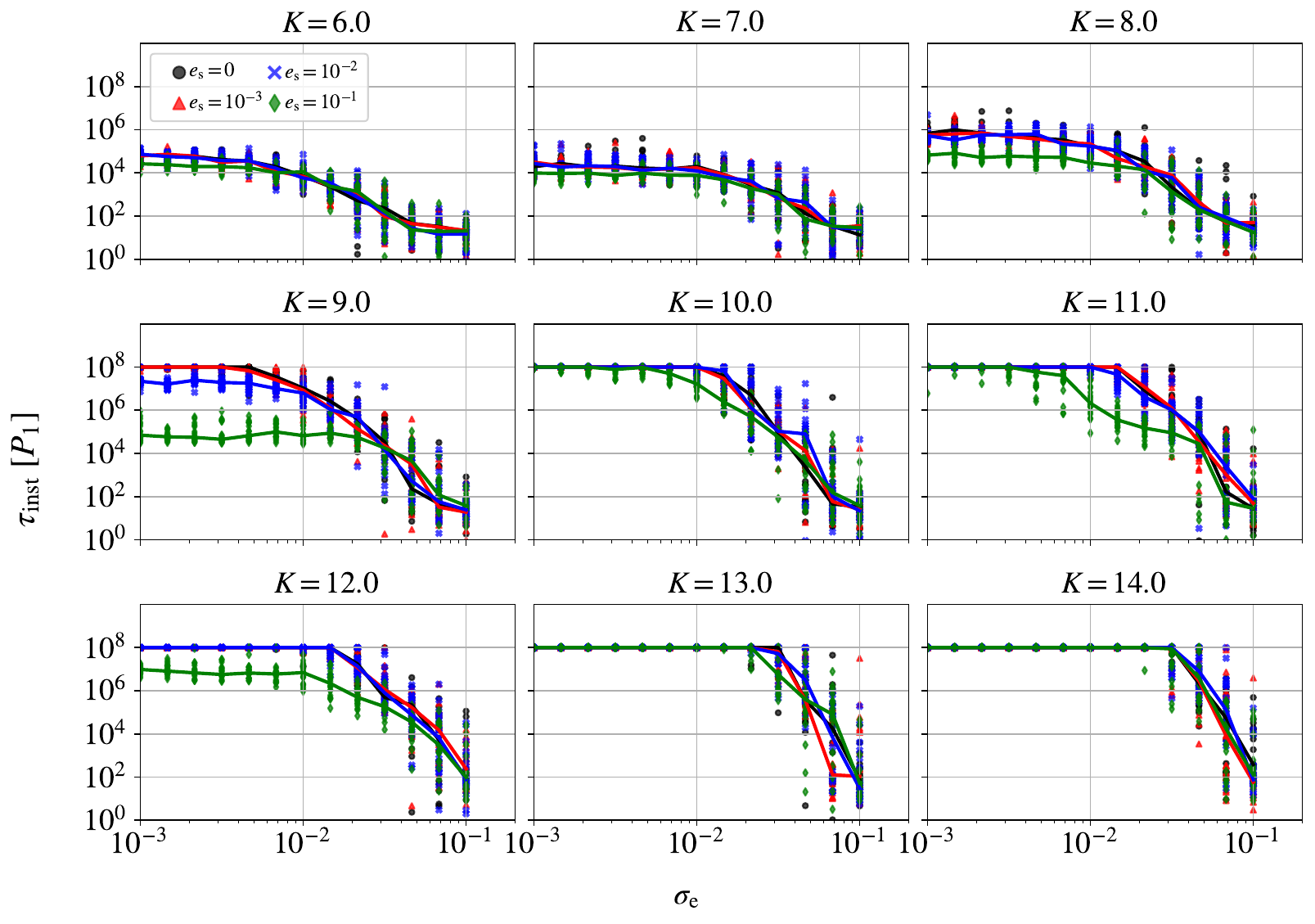}}
    \caption{Instability timescales $\tau_{\mathrm{inst}}$ as a function of dispersion eccentricity $\sigma_{\mathrm{e}}$ for different orbital separation $K$. For each pair $K, \sigma_{\mathrm{e}}$, the $\tau_{\mathrm{inst}}$ for different systematic eccentricities $e_{\mathrm{s}} = 0, 0.001, 0.01, 0.1$ are shown by black, red, blue, and green dots, respectively. The solid lines show the median $\tau_{\mathrm{inst}}$ over each $e_{\mathrm{s}}$.}
    \label{tinist_ed_ks}
\end{figure*}

Figures \ref{dense_tinst_ks_ed} and \ref{dense_tinst_ks_es} show that systems with randomized longitudes of pericenter are more sensitive to initial eccentricities, while Figure \ref{tinst_map_es_ed} suggests that a wide range of eccentricities can be considered stable provided that $\sigma_{\mathrm{e}}$ is small enough. 
Now, a systematic survey of orbital separations and eccentricities is conducted to more clearly understand the relationship between the two eccentricity components and the instability timescales.
Figure \ref{tinist_ed_ks} shows the relationship between $\tau_{\mathrm{inst}}$ and $\sigma_{\mathrm{e}}$ for $K\in[6, 14]$ and $e_\mathrm{s}\in[0, 0.1]$. 
		
For small eccentricity dispersion $\sigma_{\mathrm{e}}<0.01$, there are small fluctuations in the instability timescales, even when considering larger $e_{\mathrm{s}}$ up to $0.1$, for most $K$. 
In this case, provided that the relative eccentricity between planets is small (i.e., $\sigma_{\mathrm{e}}\ll1$, apsidal alignment), instability timescales are comparable to those of initially circular systems, even for larger $e_{\mathrm{s}} = 0.1$, in agreement with \cite{tamayo21} and \cite{yee}.
		
For $\sigma_{\mathrm{e}}\geq0.01$, $\tau_{\mathrm{inst}}$ drastically decreases by 5 orders of magnitude for each order of magnitude in $\sigma_{\mathrm{e}}$. 
The effects of larger systematic eccentricities are discussed in subsection \ref{sec332}.

This is consistent with the interpretation that larger  $\sigma_{\mathrm{e}}$ can drive planet pairs into the resonance-overlap regime, in which dynamical evolution is essentially dictated by chaotic dynamics, and that the domains of MMR overlap are governed by the relative eccentricity $\bm{e}_{i, i+1}\propto\sigma_\mathrm{e}$.

As observed in Figure \ref{tinst_map_es_ed}, our results show that apsidal alignment can provide dynamical protection for planetary systems even up to larger eccentricities of $e_{\mathrm{s}} = 0.1$, without fundamentally changing the dynamical structure nor the instability timescales, provided that the eccentricity dispersion remains low.

\subsubsection{Condition for apsidal alignment}\label{sec332}
        
\begin{figure}[!ht]
    \centering\resizebox{1.0\hsize}{!}{\includegraphics[width=0.99\textwidth]{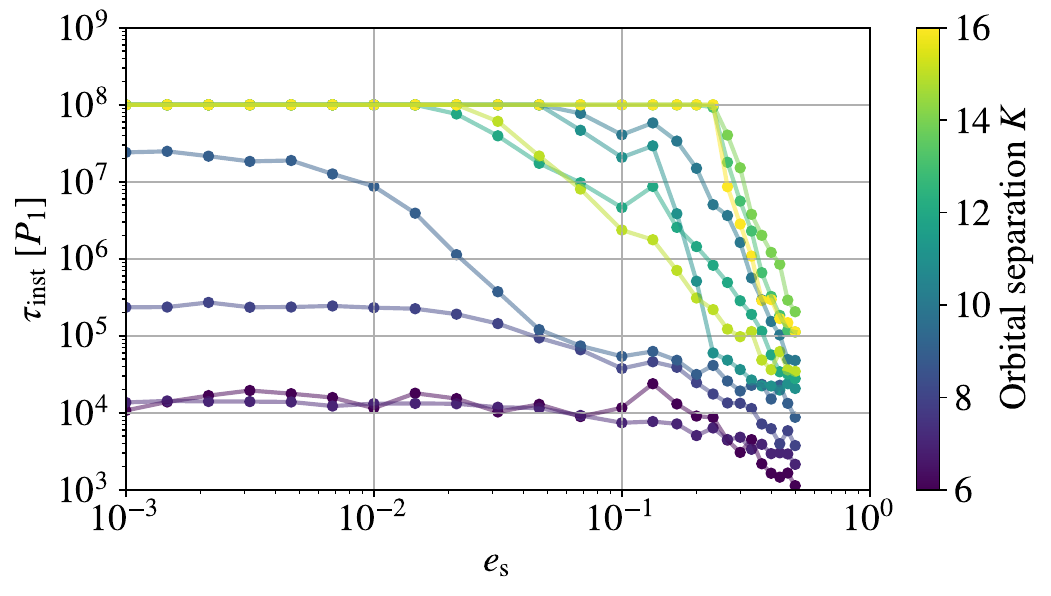}}
    \caption{Plot of $\tau_{\mathrm{inst}}$ as a function of $e_\mathrm{s}$ for different $K$. More compact systems have darker colors, whereas larger orbital separations are noted in lighter colors. To keep the plot clean, only the median values of $\tau_{\mathrm{inst}}$ over 20 runs of each pair $(e_\mathrm{s}, K)$ are plotted.}
    \label{tinist_ed_ks_apsis}
\end{figure}

To study the limitations of the apsidal alignment protection, all planets are set to be initially aligned ($e_{\mathrm{d}} = 0$), and a larger range of systematic eccentricities up to $e_{\mathrm{s}} = 0.5$ is integrated, as well as a larger orbital separation $K$. 
The median $\tau_{\mathrm{inst}}$ is plotted as a function of $e_{\mathrm{s}}$ for all different $K$ in Figure \ref{tinist_ed_ks_apsis}. 
It should be interpreted in the same way as Figure \ref{tinist_ed_ks}.

Analogously to Figure \ref{tinist_ed_ks}, a sharp decrease in $\tau_\mathrm{inst}$ can be observed for large enough $e_\mathrm{s}$.
The decrease is not as drastic as in the randomized case, though, and was empirically determined to take place at $e_\mathrm{s}^\mathrm{crit}\approx 5\sigma_\mathrm{e}^\mathrm{crit}$ (the superscript $``\mathrm{crit}"$ indicates the eccentricity beyond which $\tau_\mathrm{inst}$ decreases by one order of magnitude).

We emphasize that the condition $e_{\mathrm{s}}^{\mathrm{crit}} \cong 5\sigma_\mathrm{e}^{\mathrm{crit}}$ is empirical, and that its dynamical origin remains unclear. 
Future theoretical work will be important for better understanding this condition.

The behaviour of $\sigma_\mathrm{e}^\mathrm{crit}$ is in agreement with the analytical estimates obtained for the boundary of the MMR overlap domains (see Appendix \ref{apdxhl18} for a comparison between the analytical estimates and our values).
It can also be observed that the phase protection provided by apsidal corotation can take place for a broad range of eccentricities, provided that orbits are not in the immediate vicinity of MMR and that $\sigma_{\mathrm{e}}<\sigma^\mathrm{crit}_\mathrm{s}<e_\mathrm{s} < e_\mathrm{s}^\mathrm{crit}$.

\begin{figure}[!ht]
    \centering\resizebox{1.\hsize}{!}{\includegraphics[width=0.949\textwidth]{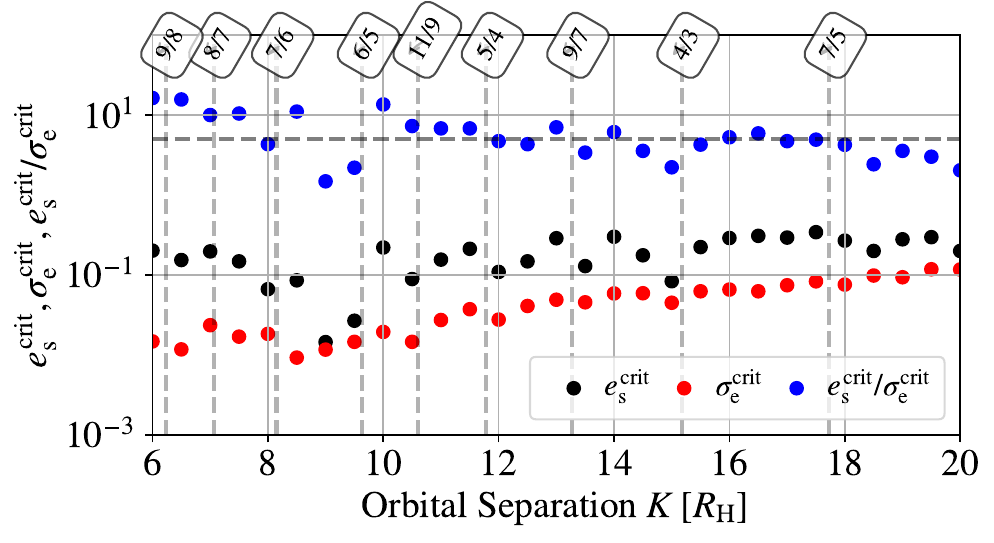}}
    \caption{Critical dispersion eccentricity $\sigma_e^{\mathrm{crit}}$ (red), critical systematic eccentricity $e_{\mathrm{s}}^{\mathrm{crit}}$ (black), and their ratio (blue) as functions of orbital separation $K$. Vertical dashed lines mark major first- and second-order MMRs, and the horizontal dashed line indicates $e_{\mathrm{s}}^{\mathrm{crit}}/\sigma_e^{\mathrm{crit}}=5$.}
    \label{ecrit_K}
\end{figure}

The values of $e_{\mathrm{s}}^{\mathrm{crit}}$, $\sigma_\mathrm{e}^{\mathrm{crit}}$ and their ratio $e_{\mathrm{s}}^{\mathrm{crit}}/\sigma_\mathrm{e}^{\mathrm{crit}}$ are plotted in Figure  \ref{ecrit_K}.

This represents an important divergence from the results obtained by \cite{lithwick} and \cite{tamayo21}, since $e_\mathrm{s}^\mathrm{crit}$ cannot be reproduced by the analytical estimates, lying at consistently larger eccentricities than predicted.

This discrepancy corroborates the presence of two distinct stability mechanisms: the instabilities observed in apsidally-aligned systems with $e_{i, i+1} = 0$ do not appear to be related to any MMR overlap, but rather to a different dynamical phenomenon.

Our framework for eccentric, apsidally aligned systems is well outside the close approximation $e_\mathrm{cross}\approx\Delta a/a$ used by \cite{lithwick, tamayo21}, given our large eccentricities.
In this scenario, according to \cite{vassiliev} orbit-intersecting criteria, $e_\mathrm{cross}$ is undefined for apsidally aligned systems, so that orbits will not intersect.

Therefore, considering that our $e_{i, i+1} = 0$ systems undergo dynamical instabilities and display chaotic behaviour for $e_{\mathrm{s}}>e_{\mathrm{s}}^\mathrm{crit}$, we further investigate the breakdown of the phase protection provided by apsidal corotation.

\subsubsection{Breakdown of apsidal alignment}\label{sec333}

Rigid-body-like precession of planetary orbits can lead to long-term stability by phase protection against close encounters, but this phase protection is not unlimited:
As the relative pericenter $\Delta\varpi$ oscillates around 0 with a non-zero amplitude, though, this phase protection can only last for as much as those orbits can oscillate around 0 without ever intersecting, meaning that the more eccentric a system is, the more restricted should be the amplitude with which $\Delta\varpi$ oscillates.

\begin{figure}[!ht]
    \centering\includegraphics[width=1.\linewidth]{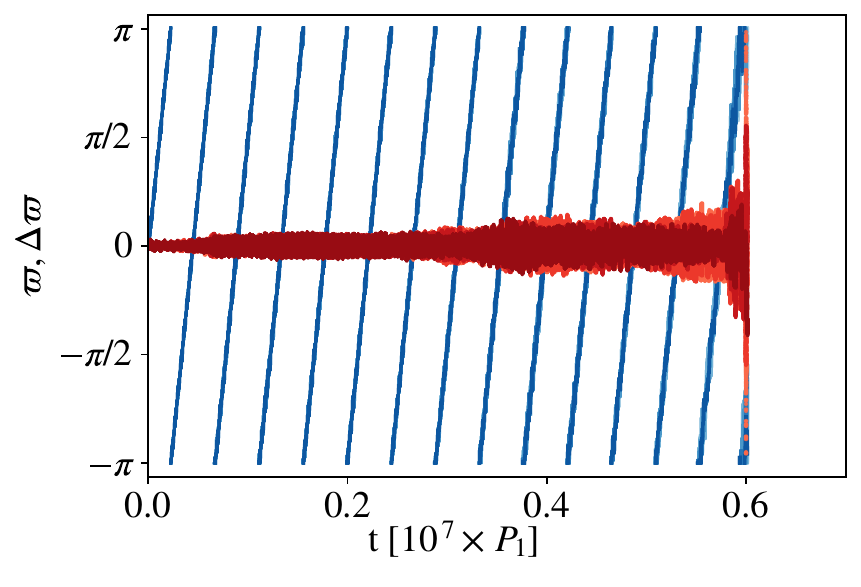}        
    \caption{Time evolution of $\varpi$ and $\Delta\varpi$ for an initially apsidally aligned system that becomes unstable, with $K = 10$, $e_\mathrm{s} = 0.1$, and $\sigma_\mathrm{e} = 0$. The loss of coherent apsidal precession near the end of the integration precedes the instability.}
    \label{orb_alg_unst}
\end{figure}

An example of an unstable aligned orbit is shown in Figure \ref{orb_alg_unst}.
As seen in panel b) of Figure \ref{dense_tinst_ks_es}, all planetary orbits precess at the same rate, rotating as a rigid body for most of the integration.
The relative pericenters $\Delta\varpi_{ij}$ oscillate with small amplitude around 0.

What distinguishes this system from the previous one is the fact that, despite being confined to low amplitudes, $\Delta\varpi$ does not display regular oscillations, but rather has a non-periodic modulation in its amplitudes, which builds up through the integration.
Shortly before $t = 5\times 10^{6}P_{1}$, the amplitude of $\Delta\varpi$ of a pair of planets suddenly increases, after which the orbits' precession decouples, leading to differential precession, thus allowing orbits to intersect. 
A physical collision occurs shortly after the breakdown of apsidal corotation.

The instability that ensues following the breakdown of apsidal corotation can be explained by \cite{lithwick} and \cite{tamayo21}: once the precession rates decouple, the planets' large $e_\mathrm{s}$ cause their relative eccentricities to increase suddenly, drastically widening the MMR.
This sudden expansion of the MMRs places the decoupled planets in the region of MMR overlap described in the previous papers.

The mechanism underlying the decoupling of apsidal corotation remains unknown, though, and should be further investigated.
A discussion of the possible physical interpretation follows.

\section{Discussion}\label{sec4}

Our numerical results indicate that the dynamical stability of compact planetary systems depends not only on the relative eccentricities but also on whether coherent apsidal precession can be maintained.

Taking into account the stability boundaries derived by \cite{lithwick} and \cite{tamayo21}, our results fully support theirs by considering that $\sigma_\mathrm{e}$ induces non-zero relative eccentricities and MMR widths for a broad range of $e_\mathrm{s}$, and a comparison between our results and the critical eccentricities derived by \cite{lithwick} can be seen in Appendix \ref{apdxhl18}. 

However, once the systematic eccentricity exceeds a critical value $e_\mathrm{s}\geq e_\mathrm{s}^\mathrm{crit}$, the coherent precession breaks down, and instability ensues even if $\sigma_\mathrm{e} = 0$, deviating from the behavior predicted by \cite{tamayo21}.

Considering the results obtained by \cite{apsidal_michtchenko} and \cite{barnes_separatrix}, one possible interpretation is that systems with large $e_\mathrm{s}$ may lie close to ``Non-linear Secular Resonance'' (NSR).
In this regime, a true secular resonance with its associated separatrix appears, which could drive chaos at larger eccentricities \citep{miga1, miga2}.

As we increase $e_\mathrm{s}$, the total Angular Momentum Deficit (AMD, \citealt{laskaramd, petitamd}) of the system increases, leading to the shrinking of the libration island associated with the NSR. 
This suggests the existence of a critical AMD$_\mathrm{crit}$, beyond which the NSR separatrix could be crossed, effectively decoupling the apsidal corotation, after which the aforementioned resonant dynamics should take place, leading to instability.
An in-depth exploration of the NSR should be conducted and will be approached in a future paper.
Future work should investigate whether a critical AMD incorporating higher-order secular theory can reproduce the newly identified stability boundary.

An empirical transitional criterion was derived between stability and instability of apsidally aligned systems $e_s^{\rm crit}\approx5\sigma_e^{\rm crit}$, but its physical origin is largely unknown. 
A deeper understanding of this threshold will require future theoretical work and is expected to shed light on the underlying dynamics.

These results suggest that future analytical stability criteria should include not only orbital eccentricities but also the dynamical state of apsidal precession. 
Such an extension would provide a more complete description of the long-term stability of compact planetary systems.

\section{Conclusions}\label{sec5}

We have shown that the dynamical stability of compact planetary systems depends not only on orbital eccentricities but also on the coherence of apsidal precession.
Systems in apsidal alignment remain nearly as stable as circular systems over a broad range of eccentricities because coherent apsidal precession suppresses orbit crossing. 
However, this protection mechanism has a finite domain of validity and breaks down above a critical systematic eccentricity $e_{\mathrm{s}}$.

We showed that the dispersion eccentricity is responsible for short-term chaos even at small $\sigma_\mathrm{e}\approx0.01$, leading planet pairs towards domains of higher-order MMR overlap \citep{wisdom_overlap, lithwick, tamayo21}.
We found that, even though apsidal alignment enhances dynamical stability \citep{chiang, malhotra_2002, zhou_aps, apsidal_michtchenko, libert}, this phase protection has a finite domain of validity, breaking down at larger $e_{\mathrm{s}}$. 

Our simulations revealed an empirical relation $e_s^{\rm crit}\approx5\sigma_e^{\rm crit}$, beyond which the instability timescales decrease sharply.
The physical origin of this relationship remains to be understood, and the validity of this relation on planetary masses should be further explored.

These results extend previous analytical stability criteria \citep{lithwick, tamayo21} by identifying a regime in which coherent apsidal precession breaks down despite the relative eccentricities being nearly zero. 

As an implication, compact planetary systems could be observed at large eccentricities, provided that they are in apsidal alignment. Such large eccentricities would otherwise place the system within the resonance overlap domain.
As apsidally aligned systems have already been detected, orbital angles, along with eccentricities, should be taken into account when assessing the stability and architecture of compact systems.

Our results demonstrate that secular apsidal dynamics constitutes an essential ingredient of the long-term stability of compact planetary systems. 
Incorporating apsidal coherence into future analytical stability criteria \citep{laskaramd, petitamd} will provide a more complete theoretical framework for understanding the architecture and evolution of planetary systems.
    
\begin{acknowledgments}
    
We thank the anonymous referee for the helpful comments. G.T.G. is funded by Mobunkagakusho scholarship from MEXT (Ministry of Education, Culture, Sports, Science and Technology of Japan).
E.K.\ is supported by JSPS KAKENHI Grants No.\ 24K00698 and 24H00017.
All work was conducted using the PCC infrastructure of the Center for Computational Astrophysics of the National Astronomical Observatory of Japan.
    
\end{acknowledgments}

\begin{contribution}
    
    
    All authors contributed equally to this work.

\end{contribution}

%

\software{REBOUND \citep{rein_mercury},
    Numpy \citep{numpy},  
    Matplotlib \citep{matplotlib}
}



\bibliography{sample8}{}
\bibliographystyle{aasjournalv7}

\appendix

\section{High-Resolution Sampling vs. Multiple Runs}\label{apdxens}

The purpose of this appendix is to demonstrate that the fitted instability relation is robust against the choice of parameter sampling, and to verify that our conclusions do not depend on the sampling strategy.
To achieve this, we compare high-resolution, single-run surveys with the lower-resolution, ensemble integrations.
		
We integrated the same systems of Subsection \ref{sec32} for evenly spaced points $4\leq K\leq8.5$ for $10^{8}P_{1}$, for different $\sigma_{\mathrm{e}} = 0.0, 0.01, 0.02, 0.05$, and fit the log-linear relationship $\log\tau_{\mathrm{inst}} = bK+c$ to the data points.
We chose $n = 1, 20, 50$ as the number of planetary systems integrated in each ensemble, and set the number of points in $K$ to $N_{K} = 1500/n$.

\begin{figure}[!ht]
            \centering
            \includegraphics[width=0.9\linewidth]{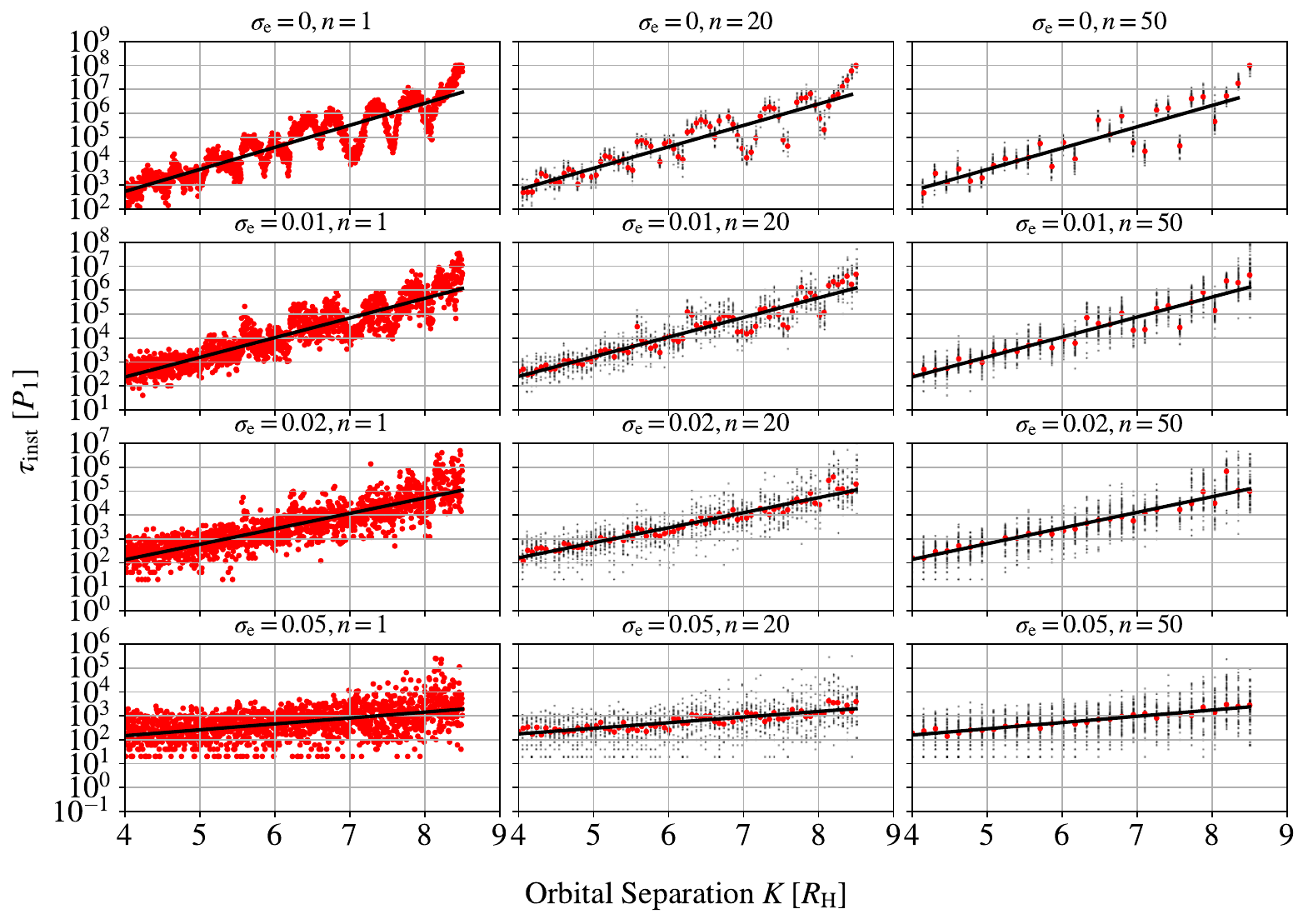}
            \caption{Relationship between instability timescales $\tau_{\mathrm{inst}}$ and orbital separation $K$ for different initial dispersion eccentricity $\sigma_{\mathrm{e}}$ and ensemble number $n$. Black curve refers to a log-linear fit of the equation $\log_{10}\tau_{\mathrm{inst}} = bK+c$. }
            \label{ens_res}
		\end{figure}

The results can be seen in Figure \ref{ens_res}, whereas the coefficients $b$, $c$ as a function of $n$ and $\sigma_{\mathrm{e}}$ and their respective errors $\sigma_{b}, \sigma_{c}$ are shown in Figure \ref{coefs_apdx} and Table \ref{tab_fit_ens}.
		
It can be seen that the fitting coefficients are consistent across sampling strategies and remain statistically stable.
This result justifies the use of high-resolution single-run surveys throughout the main text, allowing the fine dynamical structure of the phase space to be resolved without introducing a significant bias in the fitted instability relation. 
Note that the computational effort is approximately the same for all setups.

Figure \ref{coefs_apdx} summarizes the results, showing that the scatter in $b$ and $c$ is within statistical uncertainty and does not show a systematic trend with the size of the ensemble.
We therefore conclude that the choice between dense parameter-space sampling and ensemble averaging does not affect the principal conclusions of this study.
        
        \begin{minipage}[c]{.52\textwidth}
                \includegraphics[width=1\textwidth]{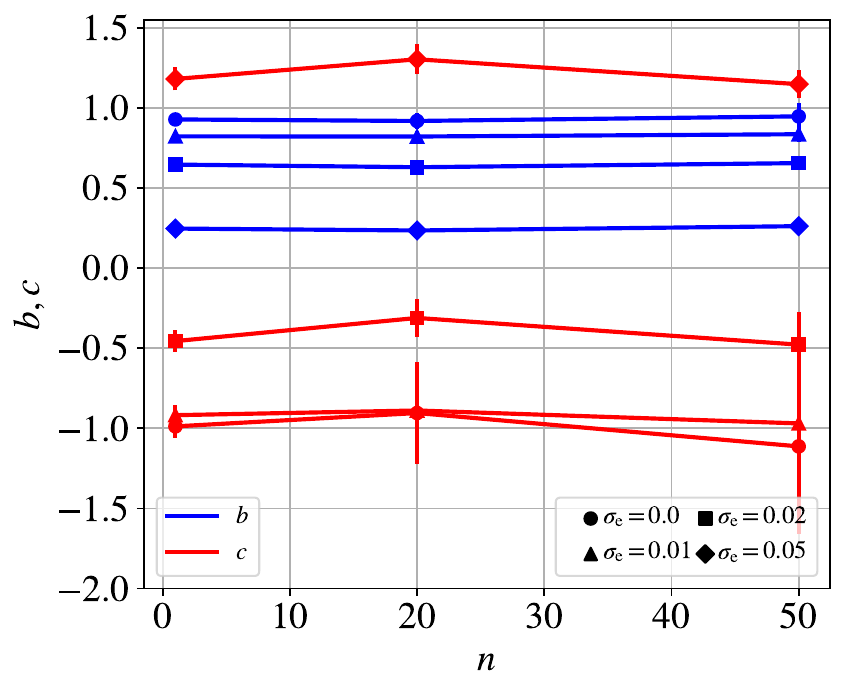}
                \captionof{figure}{\footnotesize Coefficients $b$ and $c$ of the log-linear fit $\log_{10}\tau_{\mathrm{inst}} = bK+c$ as a function of the number of simulations $n$ per initial condition for a set of initial eccentricity dispersions }
                \label{coefs_apdx}
        \end{minipage}
        \hfill
        \begin{minipage}[c]{0.44\textwidth}
            \centering
            \resizebox{\linewidth}{!}{
            \begin{tabular}{ccccccc}
                \toprule
                $\sigma_{\mathrm{e}}$ & $n$ & $b$ & $\sigma_\mathrm{b}$ & $c$ & $\sigma_\mathrm{c}$ \\
                \midrule
                \multirow{3}{*}{0}
                & 1  & 0.928 & 0.011 & -0.988 & 0.072 \\
                & 20 & 0.918 & 0.049 & -0.905 & 0.317 \\
                & 50 & 0.947 & 0.085 & -1.113 & 0.547 \\
                \midrule
                \multirow{3}{*}{0.01}
                & 1  & 0.822 & 0.010 & -0.918 & 0.061 \\
                & 20 & 0.821 & 0.032 & -0.889 & 0.203 \\
                & 50 & 0.836 & 0.048 & -0.970 & 0.305 \\
                \midrule
                \multirow{3}{*}{0.02}
                & 1  & 0.646 & 0.010 & -0.456 & 0.067 \\
                & 20 & 0.630 & 0.018 & -0.311 & 0.116 \\
                & 50 & 0.655 & 0.032 & -0.478 & 0.206 \\
                \midrule
                \multirow{3}{*}{0.05}
                & 1  & 0.247 & 0.011 & 1.181 & 0.071 \\
                & 20 & 0.235 & 0.015 & 1.303 & 0.093 \\
                & 50 & 0.262 & 0.014 & 1.148 & 0.089 \\
                \bottomrule
            \end{tabular}}
            \captionof{table}{\footnotesize Log-linear fit coefficients $b$, $c$ and their uncertainties $\sigma_\mathrm{b}$, $\sigma_\mathrm{c}$ for different values of $n$, $\sigma_{\mathrm{e}}$}
            \label{tab_fit_ens}

        \end{minipage}

\section{Impact of the Definition of Instability}\label{sectrig}

We examine whether the choice of instability criterion affects the measured instability timescales, and justify our choice of the close encounter timescale to be defined as the instability timescale $\tau_\mathrm{close} = \tau_\mathrm{inst}$.

\subsection{Effects of $e$ and $I$}

\begin{figure*}[!ht]
    \centering\resizebox{1.\hsize}{!}{\includegraphics[width=1.0\textwidth]{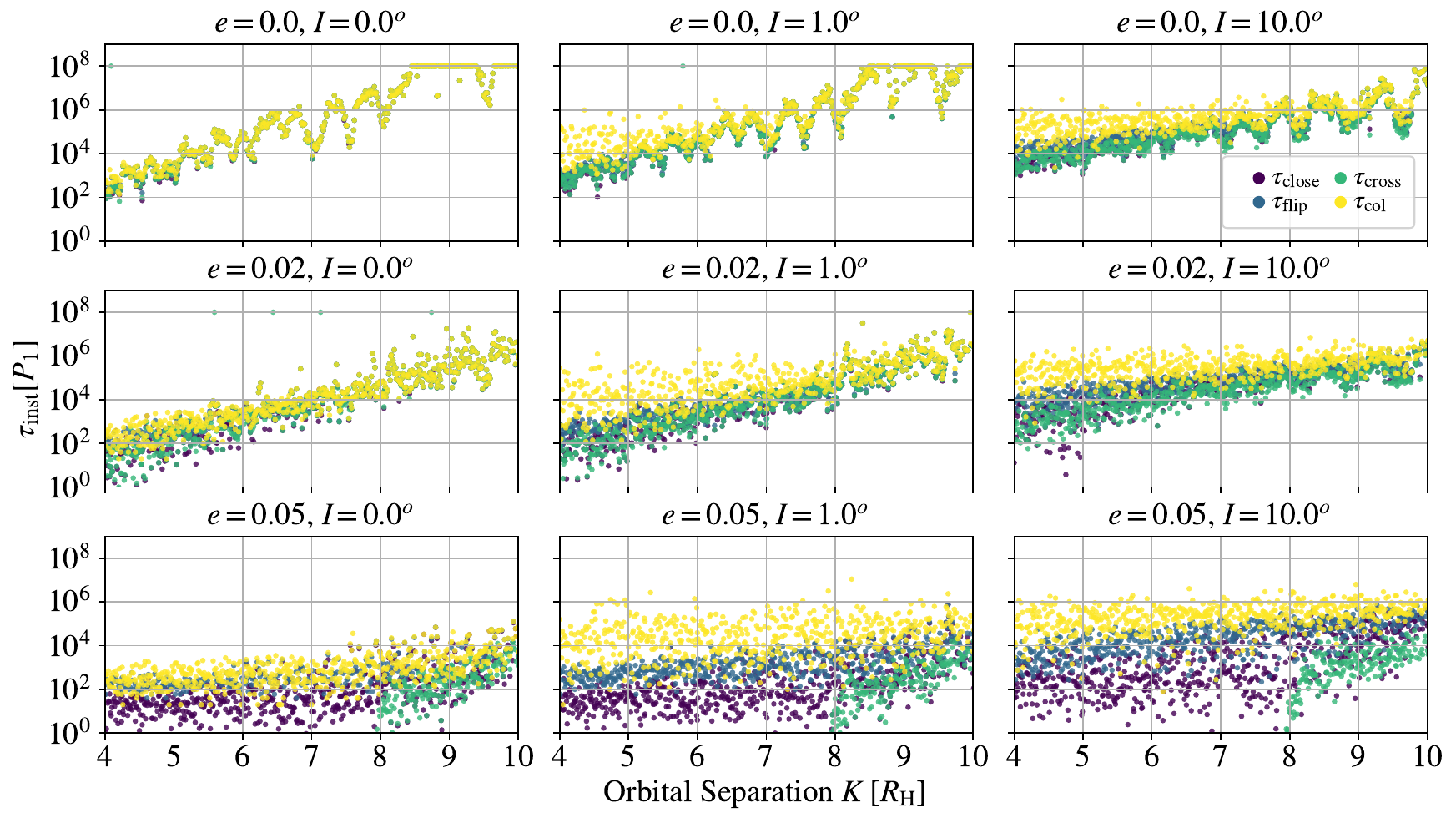}}
    \caption{Instability timescales $\tau_{\mathrm{inst}}$ as a function of orbital separation $K$ for different initial eccentricity $e$ and mutual inclination $I$. Colors indicate different definitions of instability: Close encounter timescales $\tau_{\mathrm{close}}$, orbital flip $\tau_{\mathrm{flip}}$, orbit crossing $\tau_{\mathrm{cross}}$, and physical collisions $\tau_{\mathrm{coll}}$ (black, red, blue, and green, respectively).}
    \label{tinst_ks_bench_triggers}
\end{figure*}

Figure~\ref{tinst_ks_bench_triggers} shows the instability timescale $\tau_{\mathrm{inst}}$ as a function of orbital separation $K$, calculated using four different definitions for different sets of eccentricities and inclinations. 

The top-left plot considers circular and coplanar systems. 
In this configuration, all definitions yield the same numerical results, indicating that the instability definition does not affect the simulation results. 
The high-resolution sampling in $K$ makes the dynamical structure of this configuration evident, including dips in the instability timescales around Mean Motion Resonances (MMR; \citealt{obertas17}). 

As mutual inclinations increase (from left to right), a divergence between timescales emerges.
For circular cases (three top panels), for $K\leq7$, systems can survive up to 100 times longer between the first instability and a physical collision.
As $K$ increases, the definitions again converge, indicating that the effects of inclination are not definition-dependent. The resonant structure remains clear, whilst $\tau_\mathrm{inst}$ decreases slightly.
		
Regarding the impact of eccentricities on the different instability timescales, the results are displayed from top to bottom for $e \in [0, 0.02, 0.05]$. 
Restricting to coplanar cases (three panels on the left), all but the orbit-crossing criteria converge.
This is expected as, for lower $K$ and larger $e$, adjacent planets start at $a_{i}(1+e_{i})>a_{i+1}(1-e_{i+1})$ at $t = 0$, systematically lowering the timescales for smaller $K$. 

When both initial eccentricities and inclinations are taken into account, the dynamical behaviour changes. 
Inclinations seem to counterbalance the instabilities caused by increasing eccentricities: larger $I$ allows the resonant structure to remain visible at larger $K$. 
Inclined systems can also survive longer at higher eccentricities, as can be seen in panels with $e = 0.02$, $I = 10^{o}$ and $e = 0.05$, $I = 10^{o}$.

These results are consistent with \cite{rice18}, \cite{wang_influence_2019}, and \cite{bartram}, where the authors found that inclined systems remain protected against physical collisions. 
As for coplanar and eccentric orbits, $\tau_{\mathrm{close}}$ and $\tau_{\mathrm{coll}}$ converge, indicating that collisions typically occur shortly after the first close encounter is detected.

\subsection{Survival timescales}\label{apdxtsv}

\begin{figure}[!ht]
    \centering
    \centering\resizebox{0.65\hsize}{!}{\includegraphics[width=1\textwidth]{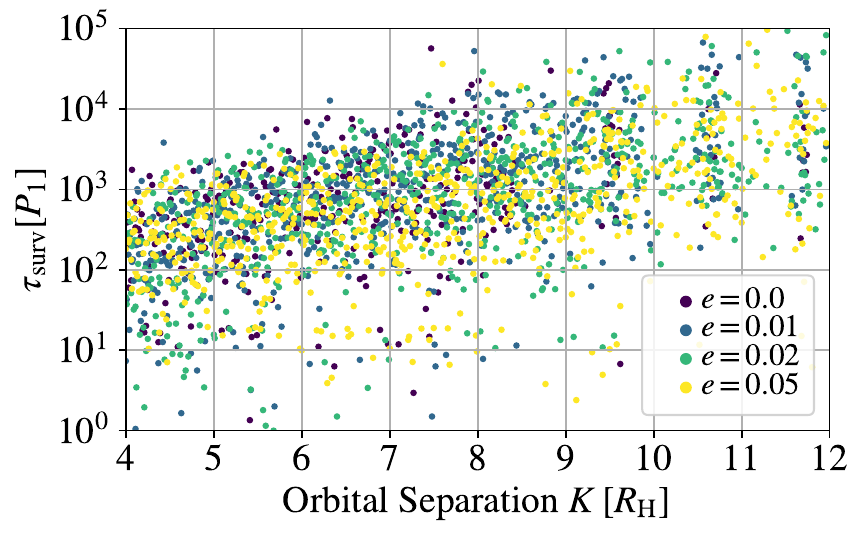}}
    \caption{Survival timescales $\tau_{\mathrm{surv}}$ as a function of initial orbital separation $K$ for different initial eccentricities plotted in Figure \ref{tinst_ks_bench_triggers}, with the addition of $e=0.01$.}
    \label{tsurv_abs}
\end{figure}

Next, we take a closer look at the survival timescales $\tau_{\mathrm{surv}}$, defined as the time elapsed between the first close encounter and the physical collision, $\tau_{\mathrm{surv}} = \mid\tau_{\mathrm{coll}}-\tau_{\mathrm{close}}\mid$.
The results obtained for four initial eccentricities (left panels of Figure \ref{tinst_ks_bench_triggers} plus $e = 0.01$) and initially coplanar systems are shown in Figure \ref{tsurv_abs}.

The survival time between the first close encounter and the first collision is generally much shorter than the instability timescale itself for the almost coplanar systems considered here, and although $\log_{10}\tau_{\mathrm{surv}}$ increases linearly with $K$ for all initial eccentricities, the slope is small. 

Having established that the close encounter time robustly measures instability for the systems considered here, we use this definition throughout the remainder of the paper to investigate the effects of apsidal alignment.

\section{Comparison with Analytic estimates of critical eccentricity}\label{apdxhl18}

We compare the original analytic estimates of critical eccentricity derived by \cite{lithwick} with the numerical results we obtained for $\sigma_\mathrm{e}^\mathrm{crit}$.

\cite{lithwick} considered the relative eccentricity between two planets $e_{12} = |e_1\exp{i\varpi_1} - e_2\exp{i\varpi_2}|$, where $i = \sqrt{-1}$, and 
analytically derived a critical value for the onset of instabilities
    \begin{equation}
        e_{12}^\mathrm{crit} = e_\mathrm{cross}\exp\left[-2.2(2\mu)^{1/3}\left(\frac{a_2}{\Delta a}\right)^{4/3}\right].
        \label{eq_hd18}
    \end{equation}

\begin{figure}[!ht]
        \centering
        \includegraphics[width=0.65\linewidth]{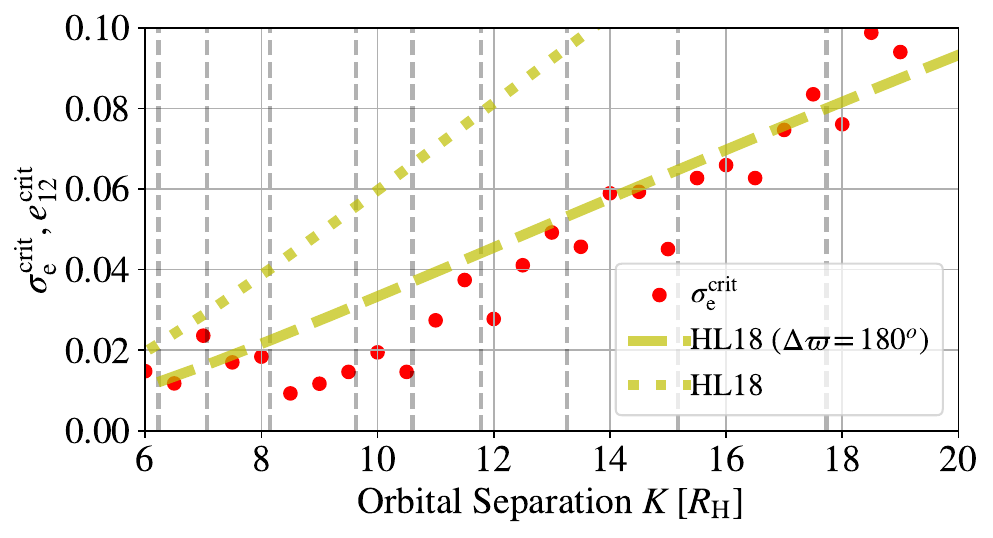}
        \caption{Same plot as shown in Figure \ref{ecrit_K}, but only considering the obtained $\sigma_\mathrm{e}^\mathrm{crit}$. The dotted yellow line refers to the critical eccentricity $e_\mathrm{crit}$ obtained by \cite{lithwick}; the dashed yellow line refers to the worst-case scenario, calculated for planet pairs initially anti-aligned $\Delta\varpi = \pi = 180^{o}$.}
        \label{ecrit_hl18}
    \end{figure}

In Figure \ref{ecrit_hl18}, the dotted line shows $e_{12}^\mathrm{crit}$ as a function of $K$.
Their estimates are expected to overestimate $e_{12}^\mathrm{crit}$, since the original criteria considered only two planets and were developed within the framework of an eccentric massless particle disturbed by an outer companion in a circular orbit, later generalized for two massive planets.

To better understand the impact of the relative pericenter $\Delta\varpi$ on critical eccentricities, we assume planetary eccentricities $e_1 \approx e_2 \approx e$, so that $e_{12} \approx 2e\mid\sin(\Delta\varpi/2)\mid$.
In this case, we can rewrite the critical relative eccentricity $e_{12}^\mathrm{crit}$ simply by expressing a critical eccentricity as a function of $\Delta\varpi$:
    \begin{equation}
        e^\mathrm{crit} \approx \frac{e_\mathrm{cross}}{2\mid\sin\left(\frac{\Delta\varpi}{2}\right)\mid}\exp\left[-2.2(2\mu)^{1/3}\left(\frac{a_2}{\Delta a}\right)^{4/3}\right].
        \label{eq_hl18}
    \end{equation}

The parameter $\sigma_\mathrm{e}$ is a proxy for the relative eccentricity, and we empirically observed $e_{ij} \approx 1.18\sigma_\mathrm{e}$ in our simulations.

The worst-case scenario (i.e., orbits start in anti-alignment $\Delta\varpi = -\pi$) is then plotted as the dashed line in Figure \ref{ecrit_hl18}.
The curve for the anti-aligned pair of planets agrees reasonably well with $\sigma_\mathrm{e}^\mathrm{crit}$, delineating the boundary for the onset of MMR overlap.
We emphasize that this result is essentially empirical, obtained from a two-planet system, and applied to a three-planet one.
We should further investigate the consistency between the line and our data points.
    
\end{document}